# Form–Substance Discrimination: Concept, Cognition, and Pedagogy

**Alexander M. Sidorkin, California State University Sacramento,** sidorkin@csus.edu

## Abstract

The skill to separate form from substance in writing has gained new prominence in the age of AI-generated content. The challenge - discriminating between fluent expression and substantive thought - constitutes a critical literacy skill for modern education. This paper examines form-substance discrimination (FSD) as an essential learning outcome for curriculum development in higher education. We analyze its cognitive foundations in fluency bias and inhibitory control, trace its evolution from composition theory concepts like "higher-order concerns," and explore how readers progress from novice acceptance of polished text to expert critical assessment. Drawing on research in cognitive psychology, composition studies, and emerging AI pedagogy, we propose practical strategies for fostering this ability through curriculum design, assessment practices, and explicit instruction. By prioritizing substance over surface in writing education, institutions can prepare students to navigate an information landscape where AI-generated content amplifies the ancient tension between style and meaning, ultimately safeguarding the value of authentic human thought in knowledge construction and communication.

## Introduction: The Challenge of Form-Substance Discrimination in Writing

The age of AI-generated text has magnified an age-old challenge in writing: distinguishing polished form from substantive content. At its core, form-substance discrimination (FSD) represents a multifaceted cognitive capacity that extends beyond mere critical reading—entailing the ability to penetrate the veneer of elegant prose to evaluate the intellectual merit of underlying ideas. This sophisticated literacy skill involves the deliberate separation of two distinct dimensions: rhetorical presentation (form) and intellectual content (substance) (McCutchen, 2011).

Fluent, error-free prose has traditionally been equated with clear thinking—"clear writing signals clear thinking," as the saying goes. This assumption held merit when only humans penned text, since producing lucid writing usually demanded genuine understanding. However, generative AI can now produce well-structured prose devoid of original thought, resulting in a new kind of "fluent emptiness" where writing sounds authoritative and coherent while masking shallow, misleading, or nonsensical content. This phenomenon reveals a deeper truth about human cognition and communication: we have long privileged the surface qualities of expression, often confusing eloquence with wisdom or accuracy with insight. The ability to craft sentences that flow smoothly represents merely one minor dimension of intellectual work, yet education systems have often rewarded formal correctness over conceptual depth. AI technology simply holds a mirror to this tendency, amplifying what was always a limitation in our approach to literacy.

The formal dimension includes stylistic elements such as sentence structure, vocabulary precision, organizational flow, and mechanical correctness—features that create an impression of sophistication and authority. The substantive dimension encompasses the quality of ideas, logical consistency, evidential foundation, and conceptual originality—elements that constitute genuine intellectual contribution regardless of presentational polish. The skilled practitioner of FSD maintains distinct mental categories for these dimensions, evaluating each independently rather than allowing surface impressions to determine overall judgment of quality (Ali, 2016).

This cognitive separation develops hierarchically. Novice readers and writers tend to perceive text holistically, with form and substance fused into a single impression of quality. As this ability develops, individuals become capable of recognizing when formally impressive text lacks substantive merit—detecting what might be termed "eloquent emptiness." At advanced levels, FSD practitioners can systematically evaluate substantive elements despite significant formal distraction, maintaining critical distance from the seductive influence of stylistic sophistication (Breuer, 2017).

In the contemporary information landscape, FSD has evolved from a specialized academic competency to an essential epistemic safeguard. The ability to distinguish between surface coherence and substantive quality now determines not merely academic success but fundamental knowledge reliability. This skill provides protection against content that exhibits formal perfection without intellectual integrity—a phenomenon increasingly common as AI text generation becomes widespread. FSD thus constitutes a fundamental reorientation of literacy toward meaning-making rather than language processing, safeguarding the value of authentic thought against the encroachment of artificial fluency (Buendgens-Kosten, 2024).

This article examines form-substance discrimination in writing, exploring its cognitive underpinnings and pedagogical implications. We survey how this skill has been discussed in composition studies and cognitive science (sometimes under different names), and how readers and writers develop the ability to separate a text's surface-level fluency from its substantive content. We then consider how this ability relates to metacognition, cognitive inhibition, and critical thinking processes. Finally, we discuss implications for teaching writing and designing curriculum and assessments in an AI-augmented era. By drawing on scholarly literature in writing pedagogy, composition theory, and cognitive research, we aim to illuminate how writing instruction can adapt to prioritize substance over surface in a landscape transformed by artificial intelligence.

## Theoretical Foundations of FSD

Form-substance discrimination has deep theoretical roots in composition studies and cognitive psychology, representing a vital literacy capacity whose importance has dramatically increased in the AI era. While scholars have long recognized the distinction between stylistic proficiency and intellectual merit in writing, this separation has recently gained new extent and new urgency.

## Theoretical Precursors

In writing instruction theory, concepts resembling FSD have been discussed extensively, though typically under different terminology. The distinction between higher-order and lower-order concerns in writing center pedagogy emphasizes prioritizing ideas, argumentation, and overall coherence over grammar, punctuation, and stylistic issues. This aligns with FSD by teaching students to focus on deeper intellectual content rather than surface-level correctness (Baaijen & Galbraith, 2018; Saeed et al., 2018).

Nancy Sommers (1980) introduced the concept that skilled writers engage in "global revision," which involves reworking fundamental ideas, arguments, or organizational logic, whereas less experienced writers typically focus only on "local revision," such as word-choice changes and sentence-level improvements. This clearly parallels FSD, emphasizing meaningful revision beyond surface polish (Saliu-Abdulahi & Hellekjær, 2020). Studies have shown that students often struggle with moving beyond local revisions, suggesting a need for pedagogical strategies that encourage them to engage with global errors and broader conceptual frameworks in writing (Saeed et al., 2018; Barkaoui, 2016).

Some writing scholars use terminology from cognitive psychology, distinguishing between "surface learning" (reproducing correct structures and facts) and "deep learning" (engaging critically with content, producing meaningful original thought). Classical rhetoric and contemporary rhetorical theory have long distinguished between rhetorical competence (effective argumentation, logical coherence, persuasive power) and mechanical competence (grammar, mechanics, stylistic correctness) (Huang & Ball, 2024; Ho, 2015). This delineation grows more complex in computer-mediated environments, where technology integration shifts traditional pedagogical boundaries (Burkhart et al., 2021).

Writing center pedagogy has long differentiated between Higher-Order Concerns (HOCs) and Lower-Order Concerns (LOCs), with the former addressing substantive elements like thesis development and logical progression, and the latter focusing on mechanical correctness and stylistic refinement (Sutherland, 2014; Almeida & Souza, 2020). This pedagogical approach acknowledges the fundamental truth that writing can be grammatically impeccable yet intellectually hollow, or conversely, mechanically flawed but conceptually rich.

What makes FSD distinct from these precursors is its transformation from a specialized capacity to an essential literacy requirement in the era of AI-generated content. Previously, form-substance discrimination was confined to advanced writing instruction and professional editing contexts—a refined skill expected mainly of journal reviewers, editors, and writing instructors. Now, large language models routinely produce content that exhibits surface-level coherence while potentially harboring logical fallacies, factual inaccuracies, or conceptual vacuity (Sato & Matsushima, 2006; Ranalli et al., 2016). This represents a profound democratization of an advanced skill, as the ability to separate polished prose from substantive thought becomes necessary for all readers. The relationship between writing quality and cognitive effort has fundamentally altered—previously, well-crafted prose normally signaled intellectual investment, with the correlation between form and substance, while imperfect, reliable enough for practical purposes. However, AI has severed this connection entirely, transforming what was once an occasional pedagogical concern into an urgent literacy need, as dramatic form-substance mismatches have evolved from relatively rare occurrences, confined to parodies or extreme cases, to a pervasive phenomenon in our information ecosystem (Wolfe et al., 2009; Buendgens-Kosten, 2024).

## Cognitive Foundations

Distinguishing surface fluency from deep substance engages a complex interplay of cognitive mechanisms that psychologists and neuroscientists have meticulously mapped. Central to this phenomenon is the mind's relationship with processing fluency—the cognitive ease with which information flows through our perceptual and interpretive faculties. Research reveals a profound bias wherein humans systematically equate processing ease with validity or value. In controlled experiments, identical essays presented in legible handwriting consistently receive more favorable evaluations than their messy counterparts, demonstrating how the mere aesthetic accessibility of information creates a positive impression that unconsciously transfers to judgments about content quality (Haling, 2022).

From cognitive psychology's standpoint, fluency bias represents a fundamental cognitive tendency wherein humans mistakenly equate processing ease with intellectual value or factual accuracy. This pervasive phenomenon operates through specific neurological mechanisms—stylistically familiar or linguistically clear texts activate well-established neural pathways, reducing cognitive demands and generating positive affective responses that become unconsciously transferred to content evaluation. This misattribution process creates what researchers term the "processing fluency heuristic"—a powerful cognitive shortcut transforming comprehension ease into presumed validity (Jacoby and Dallas, 1981).

Experimental evidence supports the assertion that aesthetic accessibility influences readers' evaluations and perceived comprehension of content presented in various formats. In controlled studies, the way information is visually presented can lead to significant differences in users' preferences and their estimates of understanding, independent of the substantive quality of the content itself. For instance, Lee and Koubek reveal that there exists a notable discrepancy between perceived usability and perceived aesthetics in user interactions with different computer-based designs. Their findings demonstrate that while high aesthetic appeal correlates positively with user satisfaction, it does not necessarily guarantee high usability, proposing that aesthetic factors can skew users'; self-assessments of comprehension (Lee & Koubek, 2010).

Research illustrates that individuals often prioritize aesthetically pleasing designs over functional ones, impacting their cognitive processing of information. This phenomenon can lead to cognitive biases, fostering overconfidence in users regarding their understanding of presented material, regardless of its accuracy. A study by Crilly et al. (2006) highlights that users'; perceptions of product functionality are heavily influenced by aesthetic appeal, suggesting that attractive design can enhance perceived usefulness and thereby mislead users about their actual comprehension of the content (Crilly et al., 2006). Similarly, research focuses on health communication has shown that the visual appeal of information can evoke strong emotional responses, leading to superficial processing rather than a thorough understanding (Chen et al., 2019). Such findings reveal a critical relationship between aesthetic factors and cognitive biases, where attractive presentations can result in users overestimating their grasp of the material (Chen et al., 2019).

The emergence of large language models dramatically amplifies this cognitive vulnerability, as AI systems now generate textual outputs specifically optimized for the surface qualities known to trigger processing fluency effects—creating unprecedented challenges for educational institutions as polished rhetorical presentation becomes completely dissociated from genuine intellectual insight or factual reliability (Buendgens-Kosten, 2024).

The implications of this cognitive architecture are particularly salient in the context of AI-generated text, which can achieve unprecedented levels of surface coherence while potentially lacking semantic foundation. Effective FSD therefore requires engaging higher-order metacognitive processes that monitor and regulate our initial responses. Proficient critical readers systematically deploy metacognitive strategies—questioning assumptions, evaluating logical consistency, and scrutinizing evidential support—even when confronted with convincingly articulated text. Conversely, less developed readers often fail to detect inconsistencies or conceptual lacunae when masked by superficial coherence, accepting rhetorical fluidity as a proxy for substantive rigor (Ozyeter, 2023).

Equally significant in FSD is the role of inhibitory control—the cognitive ability to suppress automatic responses. This executive function allows readers to override their instinctive positive reaction to fluent text, creating a crucial evaluative space between perception and judgment. The psychological parallel to the classic Stroop task is illuminating: just as identifying the ink color of a word requires inhibiting the automatic reading response, evaluating content quality demands suppressing the "this sounds good" heuristic to enable deeper analysis (Şeker, 2016).

This inhibitory process resembles what experienced manuscript reviewers develop after years of evaluating submissions. A senior academic journal editor learns to consciously override the initial impression created by elegant prose and sophisticated vocabulary, focusing instead on the substantive contribution beneath the stylistic veneer. Research by Pennycook and Rand (2019) demonstrates that our cognitive systems naturally equate processing fluency with truthfulness and intellectual merit—a bias that requires active suppression. When readers encounter well-crafted AI text, they must inhibit multiple automatic responses: the professional appearance of formatting, the reassuring rhythm of balanced paragraphs, and the cognitive ease of processing familiar rhetorical structures.

This inhibition requires deliberate recalibration of perceptual thresholds. Just as editors develop immunity to being impressed by mere linguistic sophistication, readers must cultivate a healthy skepticism toward the surface features that AI systems excel at reproducing. Doing so creates essential cognitive space between initial perception and final judgment—a buffer zone where critical evaluation can occur despite the powerful influence of stylistic fluency. This explains why maintaining critical distance from persuasively written text feels effortful, particularly for novice readers who have not yet developed the necessary neural pathways for automatic inhibition of surface-level impressions.

This inhibitory process aligns with dual-process models of cognition exemplified in Kahneman's System 1 and System 2 framework (Kakhneman 2011). The rapid, intuitive judgment that equates well-crafted prose with intellectual merit (System 1) must be subjected to deliberative, analytical scrutiny (System 2). Research on "pseudo-profound bullshit" detection provides empirical support for this model: individuals with stronger analytical cognitive reflection demonstrate enhanced resistance to meaningless but syntactically sophisticated statements. In experimental settings, participants frequently attributed profundity to semantically vacuous phrases like "Hidden meaning transforms unparalleled abstract beauty," revealing how syntactic coherence can create an illusion of substantive meaning. Those who successfully identified such content as meaningless demonstrated superior cognitive inhibition and skeptical disposition—traits associated with executive functioning and critical thinking orientation (Pennycook et al. 2015).

A crucial dimension of this cognitive analysis involves recognizing the fundamental divergence between human and artificial meaning-construction processes. Human comprehension is fundamentally embodied and contextual—we derive meaning through lived experience, social knowledge, and real-world interactions. AI text generation, by contrast, involves probabilistic pattern matching without semantic grounding or phenomenological understanding. Recognizing this distinction can enhance FSD by reminding readers that linguistic sophistication in machine-generated text provides no guarantee of factual accuracy, logical coherence, or epistemic value.

The practical application of this understanding involves conscious cognitive de-biasing—replacing the intuitive "sounds good, therefore is good" heuristic with deliberate analytical protocols. By systematically questioning textual assertions—examining their real-world validity, logical structure, and evidential foundation—readers can develop resistance to the seductive appeal of surface fluency. This represents not merely a reading skill but an epistemological orientation that privileges substantive understanding over stylistic impression, particularly crucial in navigating an information landscape increasingly populated by AI-generated content of variable quality and reliability.

## Developing FSD: Metacognition and Critical Thinking

The journey from novice to expert in form-substance discrimination follows a transformative cognitive arc that fundamentally reorganizes how individuals conceptualize and evaluate written expression. This developmental progression begins with what cognitive researchers have termed "surface credibility bias"—an intuitive tendency among early-stage writers and readers to equate textual quality primarily with mechanical correctness and formal compliance. Empirical studies reveal that first-year college students frequently channel their revision efforts toward sentence-level modifications and cosmetic adjustments, operating under the assumption that meaning remains static across drafts rather than viewing it as malleable and evolving (Leggette et al., 2015). This orientation toward form over substance manifests in revision practices that prioritize linguistic refinement while leaving conceptual architecture largely unexamined.

### Developmental Trajectory and Pedagogical Interventions

The catalyst for advancement beyond this initial stage often emerges through structured pedagogical interventions that deliberately juxtapose contrasting exemplars—exposing students to texts that exhibit divergent relationships between stylistic sophistication and intellectual substance. When presented with a meticulously crafted but conceptually hollow essay alongside a structurally flawed but intellectually profound counterpart, learners begin to develop more nuanced evaluative frameworks (Chou, 2017). This experiential discovery—that eloquence can mask vacuity while stylistic imperfection can house genuine insight—constitutes a pivotal developmental milestone. Educational research indicates that such comparative analysis, facilitated through guided discussions and collaborative peer review, helps students transcend purely aesthetic evaluation criteria and develop more sophisticated conceptual taxonomies for assessing written work (Hudd et al., 2012).

As this developmental trajectory advances, learners begin to internalize specific analytical strategies that systematically disentangle form from substance. These methodologies include conceptual mapping (extracting and visually representing a text's logical structure), propositional paraphrasing (restating key assertions in alternative language to test comprehensibility), and evidential analysis (distinguishing between claims and their supporting warrants). Such practices enable readers to penetrate rhetorical veneer and assess the underlying intellectual architecture of a text—revealing whether smooth transitions mask logical discontinuities or whether elegant phrasing conceals evidentiary gaps (Ramadhanti & Yanda, 2021). The implementation of these analytical protocols transforms reading from passive consumption into active interrogation, characterized by critical questions that probe logical coherence, evidential sufficiency, and conceptual originality.

The culmination of this developmental sequence manifests in what might be termed "cognitive automaticity"—an intuitive capacity to sense a text's intellectual substance without conscious methodological application. Through repeated exposure to diverse textual examples across the quality spectrum, experienced readers develop sophisticated mental schemas that permit rapid pattern recognition of substantive quality indicators. Research demonstrates that students who regularly analyze argumentative texts progressively internalize structural markers of intellectual rigor—logical coherence, evidential substantiation, counter-argument consideration—enabling increasingly efficient discrimination between superficial rhetorical flourish and genuine analytical depth (Granello, 2001).

This developmental process exhibits significant individual variation, influenced by educational background, critical thinking disposition, and metacognitive awareness. Individuals who demonstrate heightened analytical orientation and reflective tendencies typically develop FSD abilities more rapidly and robustly. Crucially, however, research indicates that these capacities remain responsive to pedagogical intervention across developmental stages. Educational approaches that explicitly address surface credibility bias, coupled with structured practice in analytical reading strategies, yield measurable improvements in discriminatory capacity even among initially susceptible individuals.

## Metacognitive Dimensions of FSD

Form-substance discrimination intertwines intricately with established constructs of metacognition, operating at their intersection to create a distinctive cognitive capacity. Metacognition—the recursive awareness of one's own thought processes—enables writers and readers to adopt an evaluative stance toward textual interpretation and production. When writers engage metacognitively with their drafts, they transcend surface-level considerations to interrogate the substantive quality of their thinking: "Does my argument possess internal coherence beyond its rhetorical presentation? Have I generated genuine insight or merely assembled plausible-sounding assertions?" This metacognitive vigilance aligns with what educational theorists identify as self-regulatory behavior in composition practice (Negretti, 2012; Desautel, 2009).

Research demonstrates that students who develop robust metacognitive frameworks can systematically distinguish between the impression of quality created by stylistic fluency and the actual intellectual merit of their work. Some pedagogical approaches explicitly foster this awareness through structured reflective annotations, where students must articulate their reasoning processes and identify areas of conceptual uncertainty rather than merely polishing linguistic presentation (Taczak & Robertson, 2017). These metacognitive interventions strengthen students' capacity to differentiate between perceived comprehension (often influenced by processing fluency) and genuine understanding—a distinction whose importance magnifies in contexts where AI-generated content can present surface coherence without conceptual depth.

## Critical Thinking and FSD Integration

Critical thinking—characterized by purposeful, reflective analysis of information and arguments—constitutes the methodological foundation of FSD. The critical thinker approaches text with analytical precision, looking beyond rhetorical flourishes to examine the underlying logical structure, evidential support, and conceptual originality. From the perspective of rhetorical theory, this orientation prioritizes logos (logical reasoning and evidential validity) over the potentially distracting influences of ethos (perceived authorial credibility) and pathos (emotional resonance)—a particularly crucial distinction when evaluating machine-generated content that can mimic authoritative tone without possessing authentic expertise (Abrami et al., 2015; Lai, 2011).

This connection between critical thinking and FSD manifests in specific evaluative practices: the systematic identification of claims and their supporting evidence, the detection of logical fallacies despite eloquent presentation, and the recognition of conceptual originality amidst stylistic conventionality. Composition researchers have documented how skilled readers maintain critical distance from texts, suspending judgment until they have thoroughly examined the relationship between rhetorical presentation and substantive merit—a stance increasingly necessary in navigating information environments where surface coherence no longer reliably signals intellectual rigor (Marin & Halpern, 2011).

The emerging discourse on "bullshit detection"—to employ philosopher Harry Frankfurt's deliberately provocative terminology—provides a compelling contemporary lens through which to examine FSD. Frankfurt's philosophical analysis distinguishes between lying (deliberately stating falsehoods) and "bullshitting" (speaking without genuine concern for truth, prioritizing impression over accuracy). Empirical research by Pennycook and colleagues has operationalized this concept, examining why certain individuals attribute profundity to vacuous but linguistically sophisticated statements (Pennycook et al., 2015). Their findings reveal that analytical reasoning capacity and skeptical disposition significantly predict resistance to pseudo-profound content—precisely the cognitive characteristics that underlie effective FSD.

These empirical connections between critical thinking and "bullshit detection" hold profound implications for writing pedagogy. If educational systems inadvertently reward linguistic sophistication over intellectual substance—what composition theorist John Warner characterizes as "privileging surface-level correctness" over authentic inquiry—they potentially undermine students' development of FSD (Warner, 2018). Contemporary writing instruction must therefore reorient assessment frameworks to prioritize conceptual development, intellectual risk-taking, and evidential reasoning over purely stylistic proficiency—a shift that becomes increasingly urgent as AI tools make surface fluency trivially achievable.

## Information Literacy and Contemporary Applications

The integration of FSD with broader information literacy frameworks offers a promising pedagogical direction. Information literacy encompasses the ability to locate, evaluate, and effectively use information—skills that become paramount in digital contexts where presentational polish (professional web design, authoritative formatting) often masks questionable content quality. Librarians and media literacy educators emphasize source evaluation methodologies that extend naturally to FSD: examining authorial credentials, assessing evidential foundations, and scrutinizing logical coherence rather than accepting authoritative presentation uncritically (Jolley, 2018; Mackey & Jacobson, 2011).

This expanding conception of literacy represents a necessary response to an information ecosystem increasingly characterized by synthetic content generation. When AI systems can produce plausible-sounding text without human comprehension, traditional markers of quality—grammatical correctness, stylistic fluency, organizational coherence—lose their reliability as proxies for intellectual substance. The educational imperative thus shifts toward cultivating students' capacity to penetrate these surface features and evaluate content on substantive merits—precisely the skill that FSD represents.

The contemporary information landscape—characterized by algorithmic content generation and industrial-scale misinformation—elevates FSD from an academic skill to an essential epistemic safeguard. As artificial intelligence systems produce increasingly sophisticated textual outputs that simulate intellectual substance while potentially lacking genuine insight, the ability to penetrate stylistic veneer becomes crucial not merely for academic assessment but for fundamental knowledge construction. The developmental trajectory outlined above thus represents not simply an educational progression but an epistemological maturation—a journey toward intellectual discernment that equips individuals to navigate an information ecosystem where surface coherence no longer reliably signals substantive validity.

Ultimately, FSD emerges not as an isolated cognitive skill but as a fundamental epistemological orientation—a metacognitive stance that remains vigilant against the seductive influence of surface coherence. By integrating this perspective into writing pedagogy, educators prepare students not merely for academic success but for intellectual discernment in an information landscape where rhetorical sophistication increasingly diverges from substantive quality.

## Teaching FSD: Strategies and Curricular Implementation

The integration of form-substance discrimination into educational frameworks represents a fundamental epistemological shift rather than a mere curricular addition. This section outlines pedagogical approaches and institutional strategies for developing this crucial literacy, incorporating both classroom techniques and broader curricular implementations that prepare students for an information landscape transformed by AI-generated content.

The widespread unauthorized use of AI tools among students has created an unprecedented educational crisis that extends far beyond academic integrity concerns. This phenomenon fundamentally disrupts traditional assessment models across virtually all disciplines, rendering many conventional assignments functionally obsolete. When students can generate essays, lab reports, problem solutions, and creative projects through AI systems, the evaluative foundation of education itself faces existential challenge. Our traditional methods of measuring learning, designed for an era where formal execution required substantive understanding, now fail to distinguish between authentic learning and technological simulation.

This disruption leaves educational institutions with minimal options beyond comprehensive pedagogical and curricular revision. Attempting to prohibit AI use through surveillance or detection tools merely initiates a technological arms race that diverts resources from actual education. The alternative—embracing a fundamental redesign of educational practices—represents not a temporary accommodation but a necessary evolution in how we conceptualize learning assessment in the algorithmic age.

FSD requires a sophisticated pedagogical architecture that develops both analytical acuity and metacognitive awareness. The following approaches offer pathways toward cultivating this essential literacy.

## Developing Perceptual Sophistication

Students must be guided through a process of perceptual retraining—learning to see beyond the seductive surface of polished prose to evaluate its intellectual architecture. This begins with comparative analysis of texts that deliberately subvert conventional correlations between form and substance. Instructors can curate "disguised pairs"—conceptually identical texts where one exhibits impeccable form but conceptual vacuity, while the other presents profound ideas in rougher language. Through repeated exposure to such juxtapositions, students develop a heightened sensitivity to the distinction between rhetorical sophistication and intellectual depth.

In my undergraduate course, I implement this approach through structured demonstrations of well-written texts that are fundamentally void of original ideas. The pedagogical sequence unfolds methodically: first, I model critical analysis of the text, identifying specific instances where eloquent phrasing conceals intellectual emptiness. I point out rhetorical patterns that create an illusion of depth—balanced sentence structures, sophisticated vocabulary, seamless transitions—while demonstrating how these mask the absence of conceptual originality or evidential foundation.

Students then practice this analytical process themselves, gradually developing a shared vocabulary for articulating the gap between form and substance. What makes this process remarkable is how the language of FSD appears genuinely novel to many students—suggesting that previous educational experiences have not explicitly equipped them with frameworks for distinguishing stylistic polish from substantive merit.

This pedagogical approach extends beyond analysis to production. In evaluating student work, I explicitly prioritize original thinking over presentational polished, often assigning separate grades for substantive contribution and formal execution. Many students report that this represents their first educational experience where genuine conceptual innovation was explicitly demanded and rewarded. The revelation that stylistic proficiency alone cannot compensate for conceptual shallowness often produces profound cognitive dissonance, challenging entrenched assumptions about what constitutes "good writing."

Particularly revealing are students' attempts to navigate this challenge through AI assistance. When confronted with the demand for original thinking, many initially resort to asking large language models to generate "original ideas." These efforts invariably fail, as AI systems produce aesthetically polished elaborations of existing knowledge rather than genuine conceptual innovation. This technological limitation becomes a powerful pedagogical tool, revealing through direct experience the fundamental distinction between algorithmic pattern-matching and authentic human insight. Students discover firsthand that while AI excels at creating an impression of intellectual sophistication, it cannot yet replicate the conceptual originality that defines substantive human thought.

This process requires a pedagogy of sustained attention and disciplined observation, where students practice the seemingly paradoxical skill of "looking past" surface features to perceive deeper structural qualities. Through consistent practice and explicit metacognitive reflection, students gradually develop perceptual sophistication—the ability to recognize when stylistic fluency masks intellectual emptiness, a critical skill for navigating today's information landscape dominated by synthetic content optimized for impression rather than insight.

## Cultivating Cognitive Inhibition

FSD demands that students develop inhibitory control—the ability to suspend their immediate positive response to fluent writing. In my classroom, this is framed as a form of cognitive discipline that requires deliberate practice and metacognitive awareness. When introducing this concept, I emphasize that our minds naturally equate smooth writing with good thinking, creating a bias we must actively work against.

Class exercises are structured to help students counteract their intuitive judgments. I typically show a beautifully written essay containing subtle logical flaws and ask students to share their initial impressions. Most respond positively, noting the elegant prose and authoritative tone. Following this, I guide them through systematic deconstruction, asking them to identify specific features that make the text seem authoritative, and then to recognize what substantive claims or evidence might be missing despite this impression.

The resulting cognitive dissonance creates powerful learning moments. Students experience resistance when discovering that something sounding good might actually lack substance. They learn to verbalize this struggle as they notice themselves wanting to believe an author because the writing flows well, even when closer examination reveals missing evidence for central claims.

Students are encouraged to suppress their initial instinctive judgment of what constitutes a "good paper." After years of educational conditioning that rewards polished presentation, many have internalized the equation of formal correctness with intellectual merit. Breaking this association requires creating classroom norms where skepticism toward rhetorical fluency becomes expected rather than exceptional.

This approach extends beyond evaluation to students' own writing process through what I call the "sloppy jotting technique"—a deliberate practice of unstructured, error-filled brainstorming. Students are encouraged to embrace messy thinking and give themselves permission to write badly. They produce rapid, unfiltered text full of typos, grammatical errors, and half-formed ideas without stopping to edit or structure their thoughts. Many initially resist this approach, having been trained that "good writing" means error-free prose from the first draft.

The results are often transformative. Students report discovering thoughts they didn't know they had when freed from worrying about how their writing looks. This liberation from formal constraints frees cognitive resources for substantive exploration. The unpolished nature of these early drafts creates psychological distance from the text that facilitates more objective evaluation of ideas independent of their presentation.

These exercises train students to recognize and resist the cognitive biases that stylistic fluency exploits. By consciously interrupting their automatic processing pathways, students develop a metacognitive buffer space—a moment of analytical pause between perception and judgment. This cultivated hesitation becomes increasingly valuable as AI systems generate increasingly persuasive content that leverages our instinctive equating of fluency with quality.

The metacognitive orientation developed through these exercises transforms not just how students process text but how they conceptualize intellectual work itself, preparing them for an information landscape where surface signals of quality grow increasingly unreliable.

## Teaching the Linguistic Fingerprints of AI

While contemporary AI-generated text often appears indistinguishable from human writing at a casual glance, closer examination reveals subtle patterns. In my classroom, I train students to recognize these "algorithmic fingerprints"—tendencies toward certain syntactic structures, stylistic tics, or rhetorical patterns characteristic of large language models.

Our class sessions include examining pairs of human-written and AI-generated texts on identical topics to identify distinctive patterns. We collaboratively catalog linguistic features that frequently appear in AI writing. The lexical preferences become particularly revealing—AI systems demonstrate distinctive word choices that create an impression of sophistication but often signal synthetic origin. Students quickly notice that certain metaphorical terms appear with unusual frequency in AI-generated content: words like delve, tapestry, symphony, intricate, nuanced, and resonates tend to appear when AI is attempting to perform intellectual depth without substantive content.

We also track other telltale markers: excessive hedging expressions, overreliance on certain transition phrases (moreover, additionally, furthermore), absence of idiosyncratic perspective, or unnaturally balanced presentation of viewpoints. Students come to understand that human writers naturally have biases and blind spots, whereas text that presents every possible perspective with perfect equanimity often signals AI generation rather than human deliberation.

By studying these patterns across multiple examples, students develop a more refined radar for synthetic prose. Many students who previously equated complex vocabulary with intellectual sophistication begin recognizing when elaborate language masks conceptual emptiness. This attunement enables them to detect subtle artificial notes that casual readers might miss. When students can confidently identify potentially machine-generated content, they naturally adopt a more critical stance toward its substantive claims, developing what might be termed "algorithmic skepticism" as a distinct critical literacy.

The pedagogical focus extends beyond merely spotting AI writing to understanding why these patterns emerge and what they reveal about the difference between synthetic fluency and genuine thought. This deeper analysis helps students recognize that AI's linguistic patterns reflect its fundamental limitations—the prioritization of stylistic coherence over substantive originality, a tendency that mirrors broader issues in human writing education that have privileged form over substance long before AI entered the scene.

## Fostering Intellectual Confidence Through Metacognitive Transparency and AI Integration

For students to effectively practice form-substance discrimination, they must develop sufficient intellectual confidence to trust their own evaluative judgments and understand how to maintain authorial agency when using AI tools. In my classroom, I address both aspects simultaneously through transparent thinking protocols and guided AI collaboration exercises.

Many students have been conditioned by educational experiences that emphasized compliance over critical thinking. They often hesitate to challenge eloquent-sounding text, doubting their own assessment when confronted with authoritative prose. Breaking this pattern requires creating classroom environments where questioning polished content becomes normalized and expected.

I model this process by performing analyses of texts in front of students, narrating my thought process aloud while examining whether a paragraph actually advances an argument or merely restates previous points in sophisticated language. We look for abundance of terms that sound intellectual but communicate little substance - words like tapestry, symphony, delve, nuanced, and intricate interplay often appear in content that prioritizes impression over insight.

When introducing AI tools, I teach students the rich prompt technique for proper integration. This approach positions AI as a writing assistant while ensuring intellectual substance comes from the student. In practice, this means students provide substantive elements—their original ideas, evidence, conceptual frameworks, and unique perspectives—while assigning AI specific formatting and stylistic work.

This approach naturally develops an intuitive understanding of the form-substance relationship. Students learn to gauge how much intellectual content is required to produce text that maintains appropriate knowledge density. When students provide only vague instructions to AI, they see firsthand how this produces content that sounds sophisticated but lacks substantive merit. In contrast, when they provide specific arguments with supporting evidence and ask AI to format them into a coherent essay, they maintain intellectual ownership while leveraging AI's stylistic abilities.

Collaborative annotation activities further reinforce these skills. Students work together to identify and label different dimensions of textual quality, distinguishing stylistic elements from substantive contributions. We develop a shared vocabulary for discussing knowledge density in writing, making explicit what expert readers intuitively recognize. This metacognitive transparency creates a scaffold that supports students' developing ability to maintain clear boundaries between form and content.

Group discussions analyzing AI-generated text prove particularly valuable. Students examine how AI has rephrased their ideas, whether it maintained their original meaning, added unnecessary flourishes, or diluted their thinking. Through these exercises, they develop a more sophisticated understanding of both human and machine contributions to writing, recognizing that genuine intellectual work cannot be outsourced.

The most powerful outcome emerges when students begin applying these principles independently—questioning polished text regardless of its source and developing confidence in their own analytical judgments. Students who were once intimidated by academic articles discover they can separate impressive language from actual ideas. This cognitive liberation represents not merely an academic skill but an epistemological stance essential for navigating an information landscape increasingly populated by content optimized for impression rather than insight.

## Curricular Implementation and Assessment Reform

The integration of FSD into educational frameworks demands a fundamental epistemological shift rather than a superficial curricular addition. Higher education institutions must reconceptualize this skill not as an isolated competency but as a foundational literacy that permeates all disciplinary boundaries. This necessitates systematic curricular transformation across the entire academic ecosystem.

This transformation requires the comprehensive revision of hundreds of written assignments across disciplines as the traditional "college paper" faces potential extinction. The five-paragraph essay, research paper, and reflection piece—staples of higher education for generations—now operate in a landscape where students can generate these forms with minimal intellectual engagement. Rather than attempting to retrofit these conventional assignments with detection mechanisms or prohibitive policies, institutions must orchestrate a massive shift toward authentic assessment that mirrors real-world intellectual work.

Authentic assessment situates learning activities within contexts that replicate professional practice where AI collaboration is increasingly normalized. Instead of asking students to produce isolated texts demonstrating mastery in artificial settings, assignments should engage them in complex problem-solving scenarios where they must integrate AI assistance while maintaining intellectual agency. Literature students might curate annotated collections that combine AI-generated analyses with personal insights, explicitly distinguishing algorithmic pattern-matching from original interpretation. Science students could design experimental protocols using AI assistance while critically evaluating the limitations of machine-suggested methodologies.

FSD should be embedded throughout all courses that involve communication. Writing-intensive courses across disciplines need redesigned assignments that explicitly require and assess students' ability to distinguish form from substance. These assignments should deliberately incorporate AI tools rather than attempting to prohibit them, creating authentic learning scenarios where students practice human-AI collaboration while maintaining intellectual sovereignty. For example, history courses might require students to analyze AI-generated historical narratives alongside primary sources, identifying where synthetic fluency masks factual inaccuracies or interpretive limitations.

This curricular integration requires explicit introduction of FSD early in educational sequences. First-year seminar programs provide an ideal context for introducing these concepts before they become implicit expectations in upper-division coursework. General education requirements and program-level learning outcomes need revision to explicitly include form-substance discrimination—though potentially using terminology more accessible to specific disciplinary contexts, such as "evidential reasoning" in sciences or "source evaluation" in humanities.

The broader curricular shift aligns with a necessary reorganization of skill development sequences. Traditional models that emphasized procedural intermediary skills (grammar, citation formats, standard essay structures) must evolve to prioritize higher-order capacities like FSD earlier in academic programs. This realignment reflects the reality that AI tools have automated many mechanical aspects of writing, rendering them less valuable as educational priorities while elevating the importance of distinctly human capacities for discernment and originality.

Assessment frameworks require equally profound transformation. New evaluation rubrics must de-emphasize formerly foundational skills that AI systems have rendered less relevant as markers of intellectual development. Instead, these frameworks should explicitly weight conceptual originality, logical coherence, and evidential rigor above surface polish—not merely adding these as supplementary considerations but centering them as primary determinants of quality. This recalibration demands particular sensitivity to linguistic diversity—recognizing that non-native English speakers may possess sophisticated conceptual understanding despite formal imperfections, while native speakers might leverage linguistic fluency to mask conceptual gaps.

The assessment revolution necessitated by AI-generated content extends beyond procedural adjustments to encompass a fundamental reevaluation of what we measure and value. Beyond implementing safeguards against inappropriate AI assistance, institutions must develop evaluative frameworks that explicitly privilege intellectual substance over formal perfection. The era of AI-enabled content production thus paradoxically creates an opportunity to develop more equitable assessment practices that value diverse forms of intellectual contribution over narrow conventions of academic expression.

Form-substance discrimination represents just one of several emergent literacies necessitated by AI integration, suggesting the need for comprehensive curricular revision. Other essential capacities include prompt engineering, critical evaluation of AI-generated information, and metacognitive awareness of human-AI interaction patterns. Together, these competencies constitute a fundamentally new approach to preparing students for professional landscapes where AI collaboration is ubiquitous.

For FSD to become an enduring capacity rather than a classroom exercise, students must perceive its real-world relevance. Assignments must connect this skill to consequential domains where distinguishing substance from surface matters profoundly—analyzing public policy proposals, evaluating scientific claims, or assessing information related to civic decisions. When students experience how FSD enables them to make better real-world judgments—identifying substantive policy proposals beyond political rhetoric, or recognizing genuine expertise amid confident-sounding punditry—they internalize its value beyond academic assessment.

Perhaps most profoundly, FSD serves as a bulwark against epistemological corruption in contemporary information ecosystems. By cultivating students' capacity to detect and reject content optimized for persuasive impact rather than intellectual integrity, education institutions fulfill their ethical responsibility to prepare citizens for a world awash in synthetic persuasion. This skill constitutes a form of "epistemic self-defense"—equipping individuals to maintain intellectual sovereignty amid algorithmic content designed to exploit cognitive biases. The ability to distinguish rhetorical elegance from substantive insight becomes not merely an academic competency but a prerequisite for civic participation in techno-social systems where influence increasingly derives from control over content generation rather than commitment to truth.

The academic integrity implications extend beyond disclosure requirements to fundamental questions about the nature of intellectual work itself. As AI systems increasingly automate aspects of knowledge production previously considered uniquely human, educational institutions must articulate more precise distinctions between legitimate cognitive augmentation and inappropriate intellectual outsourcing. This distinction necessarily centers on substance over form—recognizing that while AI may enhance surface presentation, genuine intellectual contribution requires the integration of experience, judgment, and perspective that remains irreducibly human.

## Conclusion: Fostering Deeper Literacy

Form-substance discrimination emerges as a critical literacy for navigating an information landscape transformed by artificial intelligence. This ability—separating the surface appeal of polished prose from the substantive quality of ideas—represents not merely an academic skill but an epistemological orientation essential for knowledge construction in the 21st century. As AI systems continue to generate increasingly sophisticated text that simulates intellectual substance while potentially lacking genuine insight, the capacity to penetrate stylistic veneer becomes crucial for students, professionals, and citizens alike.

Educational institutions bear the responsibility of cultivating this discrimination ability through explicit instruction, carefully designed assessments, and cross-disciplinary integration. By prioritizing substance over surface in teaching and evaluation, educators uphold the fundamental value of writing as a tool for communication and knowledge-building rather than an exercise in linguistic performance. This shift realigns educational priorities with the uniquely human capacities for original thought, contextual understanding, and intellectual discernment that remain beyond algorithmic simulation.

Several areas require further empirical investigation to strengthen our understanding of FSD as both concept and practice. Research should examine how this ability develops across different educational levels and disciplinary contexts, identifying potential developmental stages and effective pedagogical interventions. Studies might explore whether certain cognitive dispositions or backgrounds predispose individuals toward stronger discrimination abilities, and how these capacities can be cultivated across diverse student populations. Additionally, research should investigate specific linguistic and structural markers that characterize substantively empty yet stylistically polished text, potentially developing more precise analytical frameworks for identifying form-substance mismatches.

As we navigate the transition to an AI-augmented educational landscape, FSD stands as a bridge between traditional academic values and emerging technological realities—ensuring that as the tools of knowledge production evolve, our commitment to intellectual substance remains unwavering.

## Declaration of generative AI and AI-assisted technologies in the writing process.

During the preparation of this work the author(s) used Claude, Grok, and ChatGPT to edit the text and improve readability. After using this tool/service, the author reviewed and edited the content as needed and takes full responsibility for the content of the published article.